\documentstyle[prl,aps]{revtex}
\begin{document}
% -----------------------------------------------------------------------------
\title{Dynamical Organization around Turbulent Bursts} 
\author{Fridolin Okkels\footnote{Electronic Address: okkels@nbi.dk} and
Mogens H. Jensen\footnote[4]{Electronic Address: mhjensen@nbi.dk}}
\address{Niels Bohr Institute 
and Center for Chaos and Turbulence Studies, 
Blegdamsvej 17, DK-2100 {\O}, Denmark }
\date{October 5, 1997}
\maketitle
% -----------------------------------------------------------------------------
\begin{abstract}
The detailed dynamics around intermittency bursts is investigated in 
turbulent shell models. We observe that the amplitude of the high
wave number velocity
modes vanishes before each burst, meaning that the fixed point
in zero and not the Kolmogorov fixed point determines the intermittency.
The phases of the field
organize during the burst, and after a burst the field oscillates
back to the laminar level. 
We explain this behavior from the variations in the values 
of the dissipation and the
advection around the zero fixed point. 
\end{abstract}
% -----------------------------------------------------------------------------

One of the most fundamental problems in turbulence research is
the understanding of intermittency effects \cite{Frisch}. In
fully developed turbulent flows laminar quiescent periods are
interrupted by strong intermittency bursts of
high energy dissipation. It is well-known both from a number of
experiments and numerical simulations that intermittency effects
cause corrections to the classical Kolmogorov theory \cite{Kol}
when the structure functions of the velocity field are statistically
averaged over space and/or time \cite{Frisch}. On the other hand
very little is known about the particular structure of intermittency,
as for instance the shape of and the behavior around a specific
intermittent burst. It is the purpose of this letter to present 
a detailed investigation of the behavior of the velocity field
before and after a burst takes place. We observe a consistent picture
in which the velocity gradients, over a small scale, always
becomes organized and vanish just before the energy burst sets in 
by an ``explosion'' in the field. This is like a ``calm before the storm''
and one can draw an analogy to depinning charge-density-waves which
phase-organizes just before they slip \cite{CDW} or
self-organized-critical systems in general where bursts or ``avalanches''
of all sizes can be triggered even by the slightest perturbation \cite{SOC}.
We draw our conclusions from investigations of shell models for
turbulence, which are completely deterministic systems where the 
intriguing structure of intermittency is created by the internal
chaotic dynamics. Our results indicate that the trivial fixed point
in zero and not the Kolmogorov fixed point \cite{mogens,kada2} is responsible
for the intermittency.
Shell models are formed by various truncation techniques 
of the Navier-Stokes equations and have become paradigm models for the 
study of turbulence at very high Reynolds numbers \cite{mogens}. The
mostly studied shell model is the ``GOY'' model of 
Gledzer-Ohkitani-Yamada 
\cite{kada2,Gledzer,OY,JPV,benzi,pisarenko,kada1,bif,Lohse}. 
This model yields corrections to the Kolmogorov theory \cite{JPV} in good 
agreement with experiments \cite{Anselmet,Sreenivasan,vdWater}. 

For the ``GOY'' shell model, wave-number space is divided into
$N$ separated shells each characterized by a wave-number
$k_n=\lambda^n \, k_0$ ($\lambda=2)$,  with $n=1,\cdots N$. The corresponding
amplitude of the velocity field at shell $n$ is a
complex variable $u_n$. By assuming interactions among 
nearest and next nearest neighbour shells and phase space
volume conservation one arrives at the following 
the evolution  equations \cite{OY}
\begin{equation}
\label{un}
({d\over dt}+\nu k_n^2 ) \ u_n \ = 
 i \,k_n (a_n \,   u^*_{n+1} u^*_{n+2} \, + \, \frac{b_n}{2} 
u^*_{n-1} u^*_{n+1} \, + \,
 \frac{c_n }{4} \,   u^*_{n-1} u^*_{n-2})  \ + \ f \delta_{n,4},
\end{equation}
with boundary conditions $b_1=b_N=c_1=c_2=a_{N-1}=a_N=0$.
$f$  is an external, constant forcing, here on the forth mode.

The  coefficients of the non-linear terms  must follow the relation
$a_n+b_{n+1}+c_{n+2}=0$ in order  to satisfy the conservation  of  energy,
$E = \sum_n |u_n|^2$, when $f=\nu = 0$. 
The constraints still leave a free parameter $\delta$ so that 
 one can set
$ a_n=1,\ b_n=-\delta,\ c_n=-(1-\delta)$ \cite{bif}. If 
helicity conservation is also demanded, one obtains the canonical
value $\delta = 1/2$ \cite{kada1}. The set (\ref{un}) of $N$ coupled
ordinary differential equations can be numerically integrated by
standard techniques.
In the simulations, we use the following values: $\delta =1/2,
N=19, \nu = 10^{-6}, k_0 = 2^{-5}, f = (1+i)*0.005$.

Taking a closer look at the dynamics of the GOY model in terms
of the complex field $u_{n}(t)=r_n(t)\,e^{i\theta_n(t)}$, the intermittent
bursts consist of a collection of different organizations 
of the amplitudes $r_n$
which travel with exponentially increasing speed from the lower up 
towards the higher shells where they are damped away 
by viscosity \cite{Dombre}.
Every burst in the model follow a common pattern, where the most prominent
characteristic is that the amplitudes of
the higher modes vanish just before a burst, as shown 
in Fig.1. During the attraction towards $u_n = 0$ the phases $\theta_n$
organize in period three in the shell index $n$. Just at the point of  
minimum amplitude, the phases change so that a new organization
of period three occurs during the rapid repulsion from zero. 
After the burst the modes oscillate back to the laminar level with
increasing oscillation periods (Fig.1).

In order to explain these findings we will use that the period 
three organization of the phases is present not only at bursts, but also 
during most of the evolution. To argue for this, Fig.2 shows
a long time average of the phase differences
$P_n(t) = |  \theta_n-\theta_{n-3}  |$. We observe, that the
assumption holds very well for the highest shells, i.e. that 
$\theta_{n-1}=\theta_{n+2}$, $\theta_{n-2}=\theta_{n+1}$ \cite{thesis}. 
With this assumption, the GOY model can be separated into two 
coupled ODE's controlling the evolution of the moduli 
$r_n(t)$ and the phases $\theta_n(t)$:
\begin{equation}
\label{red-pol-GOY}
\dot{r}_{n}+i\,r_{n}\dot{\theta}_{n} =
	-\nu k_{n}^2 r_{n} +i k_{n} 
	e^{-i\,S_{n}}\Big( r_{n+1} r_{n+2}
	-\frac{1}{4} r_{n-1} r_{n+1}-\frac{1}{8} r_{n-2} r_{n-1}\Big)~~. 
\end{equation}
Both sides have been multiplied by $e^{-i\theta_{n}}$ and we have
introduced the new variable $S_{n}(t)=
\sum_{j=0}^2 \theta_{n+j}(t), \quad n=1,\ldots, N-2$ with the boundary 
conditions $S_{N-2}(t)=S_{N-1}(t)=S_N(t)$. $S_n$ is the natural phase parameter
because the evolution of the model is invariant under any 
rearrangement of the phases in which the values of $S_n$ are conserved 
\cite{kada1,thesis,gat}. By separating Eq.\ref{red-pol-GOY} into real and 
imaginary part, one obtains
\begin{eqnarray}
\label{sep-GOY-re}
\dot{r}_{n} & = & -\nu k_{n}^2 r_{n} +
	\sin(S_{n})\cdot R_{n} \\
\label{sep-GOY-im}
 \dot{\theta}_{n} & = & \cos(S_{n})\cdot R_{n} / r_{n}.
\end{eqnarray} 
where
\begin{equation}
\label{Def-Kn}
R_{n} = k_{n}\left(r_{n+1}r_{n+2}-\frac{1}{4}r_{n-1}r_{n+1}
-\frac{1}{8}r_{n-2}r_{n-1}\right).
\end{equation} 
$R_{n}$ is the real valued coupling from the nearest shells
on the ${\rm n}^{th}$ shell. Combining Eq.\ref{sep-GOY-re} and 
Eq.\ref{sep-GOY-im} one eliminates the coupling from the nearest shell and 
obtains an equation for the time derivative of the phases:
\begin{equation}
\label{th-dot}
\dot{\theta}_{n}=\cot(S_{n})\left(\frac{\dot{r}_n}{r_n}+\nu k_{n}^2 \right)
\end{equation}
If we ignore the two neighbouring phases in Eq.\ref{th-dot} and replace 
$S_n$ with $\theta_n$, simple linear stability analysis of Eq.\ref{th-dot} 
gives that the phases are attracted/repelled from the fixed points
$\theta_n=\pm \frac{\pi}{2}$ depending on whether 
$\left(\frac{\dot{r}_n}{r_n}+\nu k_{n}^2 \right)$ is negative/positive. Nearly 
the same stability conditions is found for $S_n$ when the effect of the 
neighbouring phases has been included in Eq.\ref{th-dot} \cite{thesis}.
Direct measurements of the stability of the phases show excellent agreement 
with the stability predicted by the sign of 
$\left(\frac{\dot{r}_n}{r_n}+\nu k_{n}^2 \right)$. 
The only coupling from the phases on the equation of the moduli 
(Eq.\ref{sep-GOY-re}) is the factor $\sin(S_n)$ which is close to minus one
during most of the evolution. If we set $\sin(S_n)$ equal to -1, 
the effect of the phases is removed from the equation of the moduli which then 
becomes a GOY model in terms of real variables. By comparing the 
evolution of the complex and the real valued GOY model we get an estimate of 
the effect of the phases on the evolution of the complex GOY model. It turns 
out that the phases have roughly no effect on the appearance of bursts since 
the real valued model create approximately the same bursts as the complex 
model. We therefore begin by studying the bursts of the real valued model. 
The main feature of bursts is the
 attraction and repulsion of the amplitudes to $r_n =0$, which is a trivial 
but important fixed point for the model. This dynamics appears as a result of 
a balance between the viscosity term and the coupling term. In
the simplest form the real valued GOY model can be written as 
\begin{equation}
\label{rVC}
\dot{r}_n=V_n+C_n
\end{equation}
where $V_n=-\nu k_n^2 r_n$ is the viscosity term and 
$C_n=-k_n(r_{n+1}r_{n+2}-\frac{1}{4}r_{n-1}r_{n+1}-\frac{1}{8}r_{n-2}r_{n-1})$
is the coupling term, and where the forcing is neglected because we focus on
the dynamics of the high wave numbers. As $V_n$ and $r_n$ are proportional
with opposite signs, Eq.\ref{rVC} can also be written as 
$\dot{V}_n=-\nu k_n^2 (V_n + C_n)$. The values of $\dot{V}_n$
are shown in Fig.3 by a grid of arrows. 
Only the $\dot{V}_n$-field is shown by arrows as this determines the
change in the attraction (or repulsion) to (or from) the fixed point.
Furthermore, the values of $\dot{C}_n$ are not universal.
Also shown are trajectories of 
$(V_n,C_n)$ during a burst, both of the real valued model 
(solid) and the complex model (dotted)
(for the complex model the moduli are drawn). The 
dashed straight line show where the flow vanishes $(\dot{r}_n=0)$.
First we notice that the qualitative similarity between bursts in the real 
valued and complex model show the weak effect of the phases on bursts.
From the flow field in Fig.3 we see that without variations in the coupling
term, the amplitude will stabilize at the dashed line. 

Each stage of the dynamics is labelled in Fig.3. During the attraction, 
labelled A, the viscosity term and the coupling term balance each other
with a slight dominance of the viscosity term. 
Because the trajectory approaches the fixed point $r_n =0$ with decreasing
velocity and because $|V_n|$ 
and $|C_n|$ are much larger than $|\dot{r}_n|$,
{\sl the trajectory always approach $r_n = 0$ tangentially to, 
but slightly below, the diagonal $V_n = - C_n$}. 
When approaching $r_n=0$ the trajectory can only be kicked away
by variations in the coupling term, and the absence of 
these variations makes the amplitude stabilize close to zero. 
The delicate balance between $V_n$ and $C_n$ is therefore
responsible for the long lasting laminar regimes between the bursts. 

As soon as a burst approaches from the lower shells 
the coupling term becomes large and the trajectory
is forced away from $r_n = 0$ in a given direction (labelled B in Fig.3)
into a regime of positive $\dot{r}_n$. 
This direction
is not universal; by construction it depends on the value of
the amplitudes in the neighboring shells \cite{Tomas}. 
At the end of the repulsion, labelled C, 
the coupling becomes less dominant over the viscosity, 
and this is seen in 
Fig.3 as a turning towards the dashed equilibrium line. As this line is 
crossed, the amplitude reaches its maximum, after which it is again attracted 
towards zero by the dominance of the viscosity. The motion is 
highly exited by the burst and the amplitude oscillates around the line. 
The oscillations are damped away and the amplitude settles again close to the 
dashed line where the picture is now repeated again beginning at A.

From the above scenario it is possible to give an explanation to why bursts 
are created in the GOY model: The amplitude gets trapped at the fixed point 
in zero and can only be released when a burst arrives from the lower shells.
As soon as a weak burst is created at the low shells, it continues all the way
to the highest shells because the stability of higher shells are changed by
the approaching burst. {\sl The intermittency 
is created by a ``domino'' effect 
through the shells.} The highest and the lowest shells evolve differently 
because the effect of the viscosity reduces towards the low shells. This 
reduces the attraction of the amplitudes towards zero, which
makes it less possible for bursts to occur and gives instead a slow random 
walk dynamics with a Gaussian statistics. The low shells therefore produce slow 
random perturbations which propagates up through the shells and
release bursts at large shells (small scale).

In conclusion, we have described the mechanism of the creation of intermittent
bursts in the GOY model.
The results show that the creation of
a burst is determined by a delicate balance between the viscosity
and advection terms. We therefore believe that a similar scenario
might be present in other intermittent, turbulent systems and
also in experiments. Our main observation is that a burst is associated
with a ``fingerprint'': The amplitudes of the high wavenumber modes
vanish before the burst. An experimental time signal, say from
hot wire measurements, might indeed show similar characteristica.
We are in the process of investigating this using
wavelet analysis around the bursts. Similar
work in this direction has also been done recently
by Camussi and Guj \cite{camussi}.

We are grateful to P. Bak, T. Bohr, S. Ciliberto, T. Dombre, K. Hansen, 
J. Kockelkoren, G. Zocchi
for discussions and suggestions.

% --------------------------------------------------------------------
% BIBLIOGRAPHY
% --------------------------------------------------------------------

% --------------------------------------------------------------------
% FIGURE CAPTIONS
% --------------------------------------------------------------------
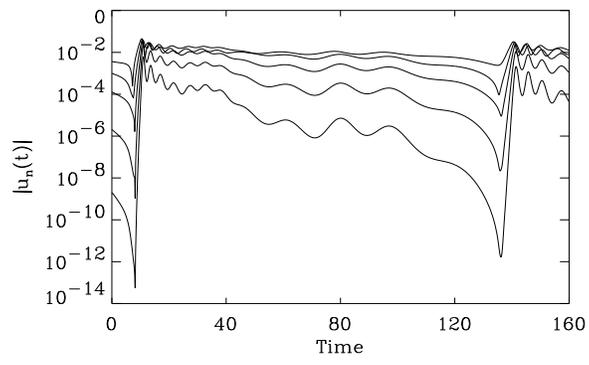
\begin{figure}
\caption[x]{
The temporal evolution (Eq.\ref{un}) of the logarithmic of the modulus of modes 
corresponding to the highest 
shells ($r_{15},\ldots,r_{19}$) for a time-span between two bursts. 
The uppermost curve corresponds to shell $15$, while the lowest to shell $19$.
The parameters for the numerical integration are listed in the text.}
\end{figure}

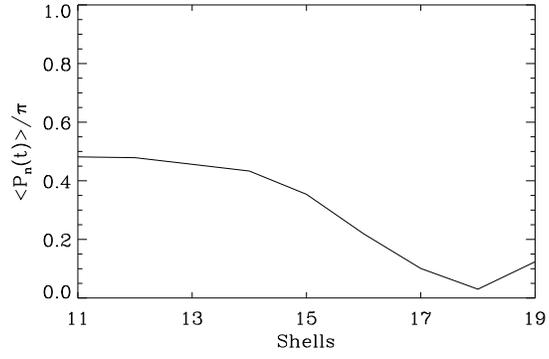
\begin{figure}
\caption[x]{
The time average of 
$P_n(t) = |  \theta_n-\theta_{n-3}  |$ normalized by $\pi$ versus the shell
number $n$. Note that the phase organization corresponding to
$P_n(t) \simeq 0$ holds very well for the highest shells. The rise
of the graph at $n = 19$ is due to a boundary effect. }
\end{figure}

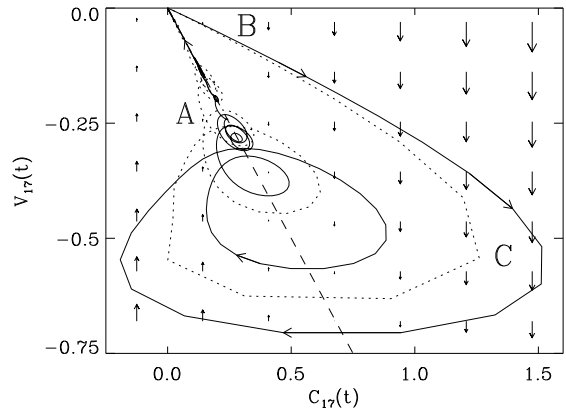
\begin{figure}
\caption[x]{
Trajectories of $V_n$ vs. $C_n$ during a burst for the real valued model,
Eq. \ref{rVC} with forcing added  
(solid curve), and the complex model, Eq. \ref{un} (dotted curve). 
On the same graph is shown the flow 
field of the viscosity $V_n$ 
visualized by arrows. 
The dashed 
line shows where $\dot{V}_n=0$. The labels represent the attraction 
to the fixed point $r_n = 0$ (A), 
the repulsion away from the fixed point (B), and 
maximum amplitude of the field (C). The arrows on the trajectories
indicate the direction of the temporal evolution.}
\end{figure}
%------------------------------------------------------------------------------
\end{document}